\documentclass [12pt,twoside]{article}           % version LATEX2E
\usepackage[left,modulo]{lineno}
\linenumbers
\usepackage{setspace}
\countdef\arxiv=201
\ifnum\arxiv=0 \input cpreb.tex \fi

\ifnum\arxiv=1

\usepackage{epsfig,times,lscape}
\usepackage[usenames]{color}

    \ifnum\count102=1

    \fi

    \ifnum\count102=2
\fi

\newcommand{\nc}{\newcommand}

     \ifnum\count101=1
\nc{\qI}[1]{\section{{#1}}}
\nc{\qA}[1]{\subsection{{#1}}}
\nc{\qun}[1]{\subsubsection{{#1}}}
\nc{\qa}[1]{\paragraph{{#1}}}

\def\qpar{\vskip 2mm plus 0.2mm minus 0.2mm}
\def\qL{\hfill \break}
     \fi 

      \ifnum\count101=2
 \nc{\qI}[1]{\parindent=0mm \vskip 8mm 
{\centerline{\LARGE \color{red}#1}}\vskip 3mm}
\nc{\qA}[1]{\vskip 2.5mm \noindent 
{{\bf\large\color{blue}  #1}} \vskip 1mm \parindent=0mm}
 \nc{\qun}[1]{\vskip 1mm \noindent {\sl #1 }\quad }

\def\qL{\hfill \break}
\def\qpar{\vskip 2mm plus 0.2mm minus 0.2mm}

      \fi
\nc{\qfoot}[1]{\footnote{{#1}}}

      \ifnum\count101=1
\def\qbu{\hfill \par \hskip 6mm $ \bullet $ \hskip 2mm}

      \fi
      \ifnum\count101=2
\def\qbu{\hfill \par \hskip 4mm $ \bullet $ \hskip 2mm}

      \fi

\def\qparr{ \vskip 1.0mm plus 0.2mm minus 0.2mm \hangindent=10mm
\hangafter=1}

     \ifnum\count101=1 
 
     \fi
     \ifnum\count101=2

  \def\qcitb#1{\noindent \hbox to 102mm{\hfill \small #1} \vskip 1mm}
      \fi

 \def\qpages#1{\count102=0{\loop\advance\count102 by 1
 \null \vfill\eject \ifnum\count102<#1 \repeat}}

\def\qn#1{\eqno \hbox{(#1)}}

\def\qv{\vskip 0.1mm plus 0.05mm minus 0.05mm}

\def\qhw{\hskip 1.5mm}
\def\qleg#1#2#3{\noindent {\bf \small #1\qhw}{\small #2\qhw}{\it \small #3}\qv }
\fi
\begin{document}
\thispagestyle{empty}

% --------------------------------------------------------------------

      % Hauts de pages et numerotation

          % Remarque: sans le \protect --> message d'erreur (ordre fragile)
\markboth{{\sl \hfill  \hfill \protect\phantom{3}}}
        {{\protect\phantom{3}\sl \hfill  \hfill}}

% -------------------------------------------------------------------
\color{yellow} 
\hrule height 20mm depth 10mm width 170mm 
\color{black}
\vskip -2.2cm 
% Titre pour la  version Phys Rev E
%\centerline{\bf \Large }
%\vskip 2mm
%\centerline{\bf \Large to populations of living organisms.}
%\vskip 2mm
%\centerline{\bf \Large First step: measuring coupling strength}
%
 \centerline{\bf \Large Is the month of Ramadan marked by}
\vskip 2mm
 \centerline{\bf \Large a reduction in the number of suicides?}
\vskip 4mm
\centerline{\large 
Bertrand M. Roehner$ ^1 $
}

\vskip 8mm
\normalsize
{\bf Abstract}\quad For Muslims the month of Ramadan is
a time of fasting but during the evenings after sunset it is
also an occasion for family and social gatherings.
Therefore,
according to the Bertillon-Durkheim conception of suicide
(that is based on the strength of social ties),
one would expect a fall in suicide rates
during Ramadan.
Is this conjecture confirmed by observation?
That is the question addressed in the present paper.
Surprisingly,
the most tricky part of the investigation was to find
reliable monthly suicide data. In the Islamic world
Turkey seems to be the
only country whose statistical institute
publishes such observations. The data reveal indeed a fall 
of about 15\% in suicide numbers during the month
of Ramadan (with respect to same-non-Ramadan months).
As the standard deviation is only 4.7\% this effect
has a high degree of significance.
This observation, along with 
the fact that other occasions of social gathering
such as Thanksgiving or Christmas are also
marked by a drop in suicides,
adds further credence to the B-D thesis.

\vskip 3mm
\centerline{\it Version of 26 July  2013}

\vskip 3mm
{\normalsize Key-words: Suicide, group effect, social interaction,
social gathering, Durkheim, Ramadan, Thanksgiving, Christmas}
\vskip 2mm

\vskip 15mm

{\normalsize 
1: Institute for Theoretical and High Energy Physics (LPTHE),
University Pierre and Marie Curie, Paris, France. \qL
Email: roehner@lpthe.jussieu.fr
}

\vfill\eject

\qI{Suicide and social ties}
Suicide is a phenomenon which often seems to frustrate 
our ``instinctive'' expectations. For instance, in the northern
hemisphere, whether in South Korea or in Turkey, the suicide
rate is lowest in November and highest in May or June.
Another surprising observation is
that suicide is not in the least affected by socio-political events. 
Thus, even first magnitude events such as the Pearl Harbor
or 9/11 attacks
had no visible effects on suicides.
For 9/11 this is true not only
at country-wide level but even
for suicide rates in New York City itself and 
whether at monthly, weekly or daily level
(Roehner 2007, p. 205-208).
\qpar

To this day
the main guideline in the understanding of suicide is still
based on the discoveries made by the sociologists
Louis-Adolphe Bertillon and Emile Durkheim
in the late 19th century. They found that married people
had lower suicide rates than bachelors or widowed
persons of same age. Moreover, they observed that among
married people, suicide rates decreased with
increasing number of children (Bertillon 1879 p. 474,
Durkheim 1897, chapter II,3). This lead them to
the hypothesis that the propensity for suicide
was determined by the number and strength of family 
bonds. So far, however, little (if any) evidence was
available to show the effect of social links
beyond the family circle. The present paper
shows that in Turkey 
during the month of Ramadan the suicide rates are about 15\%
lower than during the same time of the year not marked
by Ramadan.

\qpar
Why do we think that this observation can be accounted for
by the Bertillon-Durkheim conjecture? To see it more clearly
one must give a closer look to the characteristics of Ramadan.
During the month of Ramadan 
Muslims fast during the day from dawn
to sunset but after sunset they
break the daily fast by sharing food with those
in need and celebrate with family and friends.
This makes Ramadan a time of social gathering.
\qpar
Of course, at this point it is impossible to know whether
the fall in suicides is due to fasting,
to social gatherings or to some other facet
of Ramadan. In order to strengthen the contention that it is
indeed social gathering which is the key-factor
one must analyze other events marked by social gathering,
such as for instance
Thanksgiving in the United States. This will be done
briefly at the end of the paper.

\qI{Data}

We need to ask ourselves the following questions. (i) What suicide data 
do we need? (ii) Do such data exist? (iii) Are they reliable?
\qpar
We need monthly data from a 
country (or a region within a country) whose
population is over 90\% Islamic. Unfortunately, in most
Islamic countries the suicide rates as reported by the 
World Health Organization are very low. In Egypt, Syria, Pakistan,
Jordan, Malaysia, Indonesia, Kuwait the rate is under 2 per 100,000.
In several (if not all) of these countries  suicides are
under-reported. This is shown convincingly in the case of Malaysia
by T. Maniam (1995) and is related with the fact that 
many suspicious deaths are just 
classified as undetermined deaths.
Eventually, Turkey turned out to be the only 
mostly Muslim country for which we were able to find
monthly statistics which seem fairly sound. These data seem to exist
since 1974 but Internet availability
is limited to the period
2000-2012. Over these 13 years the month of Ramadan shifted
from December to July (every year it moves backward by some 10 days).
\qpar

Most often Ramadan overlaps two successive months but the degree
of overlapping may change widely: from one half in each month
(as in 2004 when it covered 15 October to 14 November) to
a perfect match with a calendar month (as in 2008 when
it covered 1 September to 30 September). Needless to say, when
a calendar month is a mixture of Ramadan days and non-Ramadan
days, this is not a favorable situation for the purpose
of our experiment. Therefore we restricted our analysis to those 
cases in which the number of non-Ramadan days was less or equal to 5.
As an illustration, in 2002 the Ramadan was from 6 November to
5 December which means that in November there were 5 non-Ramadan days.
In other words, based on our criterion, 2002 can be included in
our sample of favorable cases. Altogether there are $ n=6 $ favorable 
cases, namely $ F=\{2000,2002,2003,2005,2008,2011\} $.

\qI{Method}
There will be two phases in our analysis. In the first we simply
wish to get an overall view of suicide numbers during
the Ramadan periods.

%%-----------------------------------------------
%%%% Fig. MONTHLY SUICIDES IN TURKEY
\begin{figure}[htb]
\ifnum\arxiv=0 \centerline{\psfig{width=12cm,figure=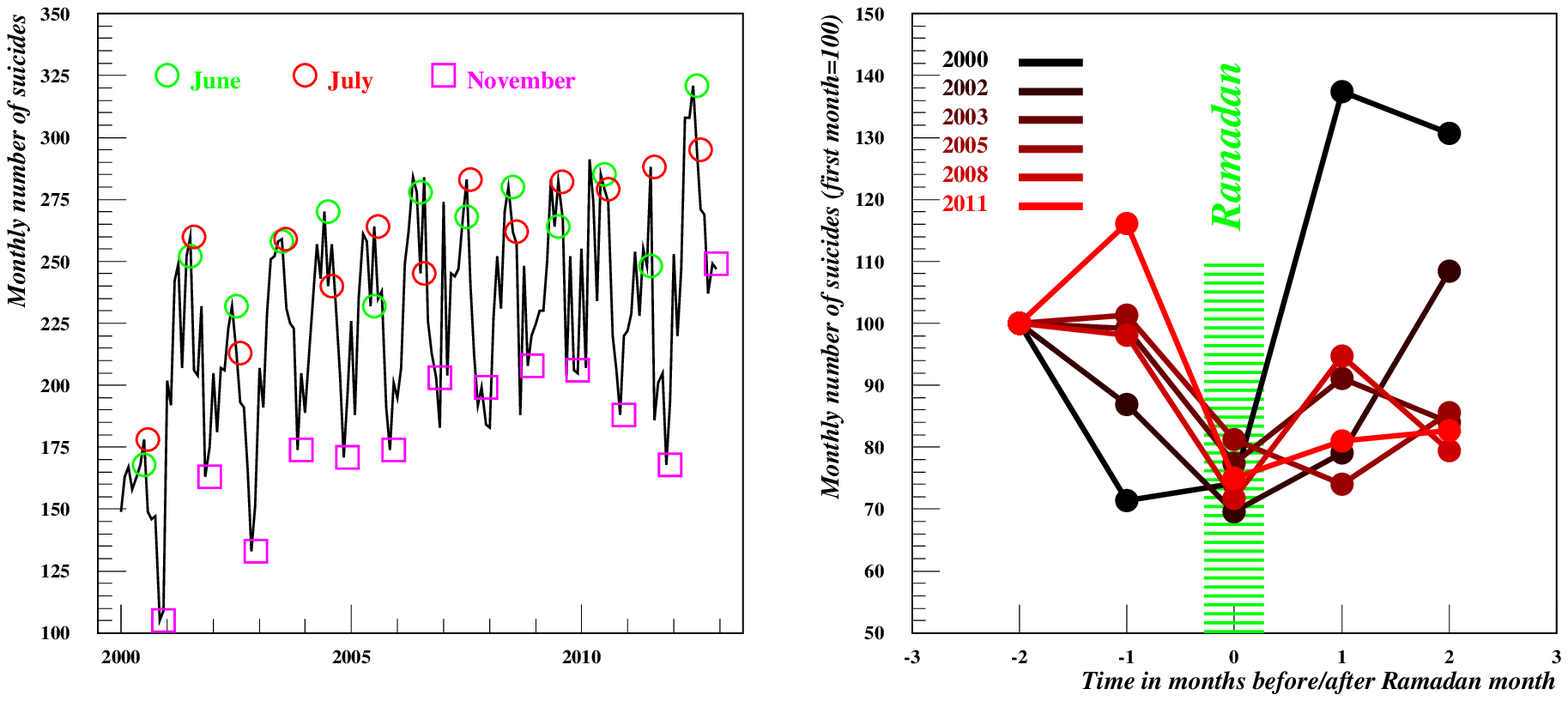}}\fi
\ifnum\arxiv=1 \centerline{\psfig{width=12cm,figure=ramadan.eps}}\fi
\qleg{Fig. 1 a,b. Left: Monthly suicides in Turkey. Right: Suicides
before, during and after the month of Ramadan.}
{Fig. 1a shows the seasonal pattern. Fig. 1b shows the
fall in suicides during the month of Ramadan.}
{Source: Suicide statistics, Turkish Statistical Institute.}
\end{figure}
%-------------------------------------------------

Fig. 1a  shows the number of suicides per month.
It reveals two features which will be of importance for our
analysis. Firstly, there is an upward trend.
Secondly, there is a strong seasonal pattern with
a maximum occurring in June or July and a minimum in November.
It is easy to get rid of the annual trend by dividing 
each year of the primary
data by their annual averages. This leads to a 
normalized time-series which is fairly stationary. 
The seasonal fluctuations are of the order of $ \pm 20\% $
which, as we will see, is of the same order of magnitude
as the Ramadan effect.
For the effect to emerge out of the background the level of noise
must be reduced by an averaging procedure over the $ n $ 
favorable cases. This will be done below in the discussion 
about statistical significance.

\qI{Results}

\qA{Graph}
Fig. 1b gives a graphical overview of the behavior of suicide
during Ramadan. It is based on the data from the normalized stationary
series. If the month of Ramadan were at the same time every year, then
Fig. 1b would have little significance. For instance, with Ramadan
in November one would anyway observe a dip around $ x=0 $ just because
of the seasonal pattern. What helps us here and makes the graph
relevant is the fact that
between 2000 and 2011 Ramadan moved from December to August. 
Therefore, the dip cannot be simply due to the seasonal pattern.

\qA{Statistical significance}
The level of significance of the dip will be estimated
through a procedure which aims at
discarding the influence of the seasonal
pattern. 
To each of the favorable years listed in Fig. 1b we associate
what will be called a reference year that is
defined in the following way.
Consider for instance 2008. Ramadan is in September. The
year 2000 will be chosen as the reference year. In 2000, Ramadan is
in December. Therefore the month of August, September 
and October 2000 will
not be affected by Ramadan and can be considered as ``reference
months''. The positivity of the difference 
$$ D=\left[S^{2008}_{\hbox{\small Aug}}- S^{2008}_{\hbox{\small Sep}} \right]
- \left[ S^{2000}_{\hbox{\small Aug}}-S^{2000}_{\hbox{\small Sep}} \right] \qn{1}$$
will test the Ramadan effect independently of the seasonal
pattern. Indeed, if the reference series decreases between
August and September (due to the seasonal pattern), then
for $ D $ nevertheless to be positive,
the Ramadan series will have to decrease
{\it more} than the reference series%
\qfoot{Can one use as reference series
the seasonal pattern itself, that is to
say the average over all years of the monthly data?
As such an average will obviously be affected by the Ramadan
effect it may seem that it cannot serve as a good reference
series. Nevertheless, it turns out that this procedure
leads to fairly similar results as the one used below.
This is certainly due to the fact that the Ramadan effect
is too weak to affect the average in any substantial way.}%
.
\qpar

For the purpose of 
estimating the statistical significance we introduce the
following notations. 
\qbu $ z_i^{(k)} $ will denote 
the variable respective to case number $ k $ of $ F $. Ramadan
month will be $ i=0 $, whereas the months before Ramadan 
and after Ramadan will be $ i=-1,1 $. 
\qbu In same way, the variable
$ a_i^{(k)} $ will denote the reference variable corresponding to case
number $ k $. By letting  $ k $ cover the set $ F $, one defines
the realizations of two random variables $ X_i $ and $ A_i $. 
\qpar
Now, in the same manner as in (1), let us define the following variables.
$$  X_i=(Z_i-A_i)/A_i,\ i=-1,0,1 \quad D_{-1}=X_{-1}-X_0,\ 
D_1=X_1-X_0 \qn{2} $$

Whether or not there is a fall in suicides during Ramadan 
will depend upon whether or not  the $ D_i $ are 
(significantly) positive.
One gets the following results for the means and standard deviations.

$$ \matrix{
D_{-1}:& \overline {D_{-1}}=5.0\%-(-11.5\%)=16.5\% & \sigma=4.75\% &
\Rightarrow & \overline {D_{-1}}/\sigma =3.4 \cr
D_1:& \overline {D_1}=-3.5\%-(-11.5\%)=8.0\% & \sigma=4.0\% &
\Rightarrow & \overline {D_1}/\sigma=2.0 \cr
} $$

If one wishes to interpret these results
in terms of significance one needs to make a specific
assumption about the probability distribution of the $ D_i $.
When (as is the case here) no information is available
regarding the probability distribution of the random variables
under consideration, it is a standard practice 
to assume that they are Gaussian%
\qfoot{In order to test this assumption one would
need some 200 data points, that is to say much more
than can ever be expected.}%
.
Under this assumption, based on a table of the
Gaussian distribution function,
one can say that $ D_{-1} $ is
positive with a degree of confidence higher than 0.9999,
whereas $ D_1 $ is positive with a degree of confidence of 0.95.
Actually, the fact that $ \overline {D_1} $ is smaller
than $ \overline {D_{-1}} $ is quite natural
because it can be expected 
that after the Ramadan shock the system
does not  return to its equilibrium (i.e. reference) 
level instantaneously. As a matter of fact,
Fig. 1b suggests that the time constant of the system is of the order
of 2 months.

\qI{Conclusion}
Is there a similar effect for Thanksgiving? The answer is yes.
Because US daily suicide statistics have been published 
since 1972, this question was already studied in the 1970s and 1980s
(see for instance Phillips and Liu 1980).
\qpar

In public holidays there are two components: (i) most of them
(e.g. Thanksgiving, Christmas, Memorial Day) are marked
by family gatherings (ii) all of them are federal holidays
which means that many persons (especially those
employed by the federal government or by state administrations)
do not have to work. \qL
Should the suicide effect be attributed to the first factor,
to the second or to both of them together? It turns out that the
suicide effect is much weaker for the Martin Luther Day
than for Christmas or Thanksgiving. Is that due to the fact
that it brings about less social gatherings or simply
to the fact that it is a day off for a smaller proportion
of people. It is difficult to know.
\qpar

In contrast, Ramadan is not a time of vacation. This narrows down
possible explanations and points at social gathering as the
key-factor, thus providing additional
backing for the Bertillon-Durkheim thesis.

\vskip 10mm

{\bf References}

\qparr
Bertillon (L.-A.) 1879: France d\'emographique. 
Article (p. 403-584)
included in: G. Masson and P. Asselin (editors):
Dictionnaire Encyclop\'edique des
Sciences M\'edicales, vol 4. Paris.

\qparr
Durkheim (E.) 1897: Le suicide: \'etude de sociologie.
F\'elix Alcan, Paris.

\qparr
Maniam (T.) 1995: Suicide and undetermined violent deaths in Malaysia,
1966-1990:
evidence for the misclassification of suicide statistics.
Asia Pacific Journal of  Public Health 8,3,181-185.

\qparr
Phillips (D.P.), Liu (J.) 1980: The frequency of suicides around
major public holidays: some surprising findings.
Suicide and Life Threatening Behavior 10,1,41-50.

\qparr
Roehner (B.M.) 2007: Driving forces in physical, biological,
and socio-economic phenomena. Cambridge University Press,
Cambridge.

\end{document}